\documentclass[conference]{IEEEtran}
\usepackage{amssymb,amsthm,amsmath,epsfig,subfigure}
\usepackage{times,fullpage}
\usepackage{cite}
\usepackage{float}
\usepackage{mathtools}
\usepackage{tabularx,booktabs}
\usepackage{algorithm,algcompatible}
\usepackage{xcolor}
\usepackage{hyperref}
\newtheorem{theorem}{Theorem}

\newtheorem{lemma}[theorem]{Lemma}

\begin{document}

\title{Truncated Polynomial Expansion-Based Detection in Massive MIMO: A Model-Driven Deep Learning Approach}

\author{

\IEEEauthorblockN{Kazem Izadinasab}
\IEEEauthorblockA{Department of Electrical and\\ Computer Engineering\\
University of Waterloo\\
Email: mkizadin@uwaterloo.ca}
\and

\IEEEauthorblockN{Ahmed Wagdy Shaban}
\IEEEauthorblockA{Department of Electrical and\\ Computer Engineering\\
University of Waterloo\\
Email: ahmed.w.shaban@uwaterloo.ca}
\and

\IEEEauthorblockN{Oussama Damen}
\IEEEauthorblockA{Department of Electrical and\\ Computer Engineering\\
University of Waterloo\\
Email: mdamen@uwaterloo.ca}
}
\maketitle

%%%%%%%%%%%%%%%%        Abstract    %%%%%%%%%%%%%%%%%%
%%%%%%%%%%%%%%%%%%%%%%%%%%%%%%%%%%%%%%%%%%%%%%%%%%%%%%

\begin{abstract}
In this paper, we propose a deep learning (DL)-based approach for efficiently computing the inverse of Hermitian matrices using truncated polynomial expansion (TPE). Our model-driven approach involves optimizing the coefficients of the TPE during an offline training procedure for a given number of TPE terms. We apply this method to signal detection in uplink massive multiple-input multiple-output (MIMO) systems, where the matrix inverse operation required by linear detectors, such as zero-forcing (ZF) and minimum mean square error (MMSE), is approximated using TPE. Our simulation results demonstrate that the proposed learned TPE-based method outperforms the conventional TPE method with optimal coefficients in terms of asymptotic convergence speed and reduces the computational complexity of the online detection stage, albeit at the expense of the offline training stage. However, the limited number of trainable parameters leads to a swift offline training process.
\end{abstract}
\vspace{2mm}

% \fi
\IEEEpeerreviewmaketitle

%%%%%%%%%%%%%%%%%%%%%%%%
%%% Introduction
%%%%%%%%%%%%%%%%%%%%%%%%
\section{Introduction}

%With the advent of powerful computing resources and also advances in big data and optimization algorithms, deep learning (DL) has revolutionized many fields, such as computer vision, natural language processing, and speech recognition. DL has been widely applied to wireless communication systems mainly because of the availability of powerful DL software libraries, its impressive computational throughput, and the ability to optimize end-to-end performance \cite{Wang2017, Dorner2018}. 

The ubiquity of powerful computing resources, coupled with the recent advances in big data and optimization algorithms, has led to a revolution in the application of deep learning (DL) in various fields, including computer vision, natural language processing, and speech recognition. DL has been widely applied
to wireless communication systems mainly because of
the availability of powerful DL software libraries, its
impressive computational throughput, and the ability to
optimize end-to-end performance to address various challenges such as channel estimation, signal detection, and resource allocation \cite{Wang2017, Dorner2018}.

In the literature, there are two approaches to the application of DL in wireless communication systems \cite{Qin2019, He2019}: The first approach is a data-driven one in which a deep neural network (DNN) is optimized using a large training data set without relying on domain knowledge and a mathematical model. The data-driven approach has been applied in many communication applications such as channel estimation \cite{Ye2018},  modulation recognition \cite{Wu2018}, and channel state information feedback\cite{Wen2018}. The performance of such a black box approach highly depends on a large amount of training data, and it is difficult to modify the topology to achieve better performances \cite{Wei2020}. An alternative to the data-driven approach is a model-driven DL, where the parameters of a model-based algorithm are optimized using DNN. Such a model-driven DL network is constructed by unfolding a well-established algorithm, also known as deep unfolding \cite{Hershey2014}. This approach allows for utilizing well-established models for physical wireless communication while taking advantage of the power of the state-of-the-art DNNs.  

The model-driven DL approach has been considered for sparse signal recovery using the approximate message passing (AMP) algorithm \cite{Borgerding2017}. In the context of signal detection for multiple-input multiple-output (MIMO), a model-driven DL approach has been utilized in \cite{Samuel2017, Samuel2019} by unfolding a projected gradient descent algorithm for maximum likelihood (ML) detection. Further, the orthogonal AMP and belief propagation algorithms have been unfolded in a model-driven DL approach, respectively, in \cite{He2020} and \cite{Tan2020}, for MIMO detection where the optimized parameters using DL improve the bit error rate (BER) performance of the original algorithms. Moreover, an iterative algorithm is proposed in \cite{Mandloi2017} and improved in \cite{Liao2020} for massive MIMO detection using the same idea. Furthermore, \cite{Wei2020} considers a model-driven DL approach using the conjugate gradient for massive MIMO detection. %  (Ahmed) I think we need to add one sentence that indicates that in all these research works the model-driven DL approach has shown significant performance improvement over the conventional ones    

In this work, motivated by the advantages of the model-driven DL approach, we propose to use it as a core machinery for optimizing the coefficients of truncated polynomial expansion (TPE) of 
 the inverse of Hermitian matrices. We employ such a model for approximating linear detectors, such as zero-forcing (ZF) and minimum mean square error (MMSE), that involve the inverse of Hermitian matrices in their calculations. In the proposed method, each term of the expansion of ZF or MMSE is viewed as one layer of the model. For a given number of TPE coefficients (or layers), the training process is performed offline using a large training data set to minimize the distance between the exact linear solution and the approximated TPE-based one. As our simulation results show, the proposed learned TPE-based detector outperforms existing TPE-based detectors in the literature \cite{Sessler2002, Izadinasab2022}. 
% Specifically,  the proposed method achieves comparable error performance to that of the exact ZF or MMSE detectors with a smaller number of layers, resulting in a lower computational complexity in the online detection stage. In this way, our proposed method strikes the right balance between the efficiency and optimality of conventional signal processing techniques (e.g., ZF in massive MIMO) and the potency of DL-based approaches. 
such that it can approach the error performance of ZF or MMSE detector with a smaller number of layers resulting in less computational complexity in the online detection stage. 

The organization of the paper is as follows: 
Section \ref{Section_System_Model} introduces the system model, while Section 
\ref{The_Proposed_Learned_TPE} provides a detailed description of the proposed model-driven DL approach for TPE. 
In Section \ref{Section_Complexity_Analysis}, we discuss the computational complexity of the proposed learned TPE-based detector for signal detection in the uplink of massive MIMO systems. Section \ref{Section_Simulation_Results} investigates the error performance of several massive MIMO systems when the proposed approach is deployed, along with the training process details. Finally, Section \ref{Section_Conclusion} concludes the paper.

{\it{Notation:}} 
We denote matrices and vectors by bold-face capital and small letters, respectively. For a given matrix ${\bf{A}}$: ${\bf{A}}^T$, ${\bf{A}}^H$, and ${\bf{A}}^{-1}$  denote the transpose, Hermitian transpose, and inverse of $\bf{A}$, respectively. 

%%%%%%%%%%%%%%%%%%%%%%%%
%%% System Model
%%%%%%%%%%%%%%%%%%%%%%%%
\section{System Model}\label{Section_System_Model}

\subsection{Truncated Polynomial Expansion (TPE)}\label{Subsection_TPE} 

According to the following lemma, the inverse of any Hermitian matrix can be expressed as a matrix polynomial. 
\begin{lemma}[\cite{Sessler2005}]\label{Lemma}
For any positive definite Hermitian matrix $\bf X$,
\begin{equation}
    {\bf X}^{-1} = \alpha ({\bf I}-({\bf I}-\alpha{\bf X}))^{-1} = \alpha \sum_{l=0}^{\infty}{({\bf I}-\alpha{\bf X})^{l}},
\end{equation}
where the second equality holds when $0 < \alpha < \frac{2}{\lambda_{max} (\bf X)}$ such that $\lambda_{max} (\bf X)$ is the largest eigenvalue of $\bf X$. The parameter $\alpha$ is referred to as the normalization factor. 
\end{lemma}
By considering a normalization factor satisfying the condition in the lemma, the eigenvalues of $({\bf I}-\alpha{\bf X})^{l}$ goes to zero as $l$ increases, and the expansion is mainly dominated by the low-order terms \cite{Muller2014}. Therefore, a truncation of the expansion using $J$ terms can be considered as an approximation for the inverse, i.e.,
\begin{equation}
    {\bf X}^{-1} \approx \alpha \sum_{l=0}^{J-1}{({\bf I}-\alpha{\bf X})^{l}},
\end{equation}
where we denote $J$ as the TPE order. Given an appropriately tuned normalization factor, the accuracy of the approximation increases as $J$ increases. However, in certain applications such as detection for the uplink of massive MIMO systems, having fewer TPE terms is desirable since it reduces the computational complexity and processing delay incurred by the iterative implementation. In the next subsection, we will describe the use of TPE for uplink massive MIMO detection.

%%%%%%%%%%%%%%%%%%%%%%%%%%
\subsection{TPE for Uplink Massive MIMO Detection}\label{Subsection_TPE}
Consider the uplink of a multiuser massive MIMO system where $K$ single-antenna user terminals (UTs) are communicating with a base station (BS) with $N$ receive antennas. The complex received signal at the BS is 
\begin{align}\label{received-sig_complex}
{\bf y}_c = {\bf H}_c{\bf x}_c+{\bf n}_c,    
\end{align}
where ${\bf x}_c \in \mathcal{X}^{K\times 1}$ is the complex transmitted signal vector whose entries are chosen from a given constellation, $\mathcal{X}$, such that ${\mathbb E} [{\bf x}_c{\bf x}_c^H] = E_x {\bf I}_K$, and ${\bf H}_c = {\frac{1}{\sqrt{N}}}[{\bf h}_1, {\bf h}_2, \dots, {\bf h}_K]  \in \mathbb{C}^{N\times K}$ is the complex channel matrix whose entries are assumed to be independent and identically distributed (i.i.d.)  $\sim {\mathcal {CN}}(0,1/N)$. Moreover, ${\bf n}_c \in \mathbb{C}^{N\times 1}$ is the complex noise vector at the BS with i.i.d. entries $\sim {\mathcal {CN}}(0,N_0)$. 
We denote $\beta =\frac{K}{N}$ as the loading factor of the massive MIMO system.

It is more convenient to consider the real-valued representation of the system model for DL implementations. Therefore, (\ref{received-sig_complex}) is rewritten as
\begin{align}\label{received-sig}
{\bf y} = {\bf H}{\bf x}+{\bf n},    
\end{align}
where 
\[
{\bf{x}}
\triangleq
\begin{bmatrix}
          \Re({\bf x}_c) \\
          \Im({\bf x}_c)            
\end{bmatrix} \in \mathbb{R}^{2K\times 1},
\]
\[
{\bf{H}}
\triangleq
\begin{bmatrix}
          \Re({\bf H}_c) & -\Im({\bf H}_c)      \\
        \Im({\bf H}_c)       & \Re({\bf H}_c)     
\end{bmatrix} \in \mathbb{R}^{2N\times 2K},
\]
\[
{\bf{y}}
\triangleq
\begin{bmatrix}
          \Re({\bf y}_c) \\
          \Im({\bf y}_c)            
\end{bmatrix} \in \mathbb{R}^{2N\times 1},
{\bf{n}}
\triangleq
\begin{bmatrix}
          \Re({\bf n}_c) \\
          \Im({\bf n}_c)            
\end{bmatrix} \in \mathbb{R}^{2N\times 1},
\]
in which $\Re(\cdot)$ and $\Im(\cdot)$ denote the real and imaginary parts, respectively.
    
Simple solutions for multiuser uplink massive MIMO are the ZF detector
\begin{align}
{{\hat {\bf x}}_{\text{ZF}}} = {\bf W}_{\text{ZF}} {\bf y} = {({\bf H}^T{\bf H})^{-1}{\bf H}^T{\bf y}},    
\end{align}
and the MMSE detector
\begin{align}
{{\hat {\bf x}}_{\text{MMSE}}} = {\bf W}_{\text{MMSE}}{\bf y} ={({\bf H}^T{\bf H}+\mu {\bf I})^{-1}{\bf H}^T{\bf y}},    
\end{align}
where $\mu= \frac{N_0}{E_x}$ is the regularization factor of the MMSE detector, and $\bf I$ is the identity matrix of appropriate dimensions. Both solutions involve the inverse of a Hermitian matrix that can be approximated using TPE. 
By using Lemma \ref{Lemma} and ${\bf X} = {\bf G}={\bf H}^T{\bf H}$, the ZF matrix can be approximated as
\begin{align} \label{W_ZF}
{\bf W}_{\text{ZF}}&={({\bf H}^T{\bf H})^{-1}}{\bf H}^T \nonumber\\
& \approx \sum_{l=0}^{J-1}{\bigg({\alpha \sum_{n=l}^{J-1}{n \choose l}(-\alpha)^{l}}\bigg)({\bf H}^T{\bf H})^l{\bf H}^T}.    
\end{align}
We denote ${\mathbf W}_{\text{TPE}}$ as the TPE detector such that 
\begin{align}\label{W_TPE}
{\mathbf W}_{\text{TPE}}= \sum_{l=0}^{J-1}{w_l\Big( {\mathbf H}^H{\mathbf H} \Big)^l {\mathbf H}^H},
\end{align}
where for ZF 
\begin{align}\label{w_l's}
w_l= {\alpha \sum_{n=l}^{J-1}{n \choose l}(-\alpha)^{l}},
\end{align}
% and for MMSE
% \begin{align}\label{}
% w_l= {\alpha \sum_{n=l}^{J-1}{n \choose l}{\big(1-\mu\big)}^{n-l}(-\alpha)^{l}},
% \end{align}
in which we refer to $w_l$'s as the coefficients of TPE. Note that the MMSE matrix can also be expanded in a similar manner. For massive MIMO systems, linear detectors using TPE are implemented iteratively through a series of matrix-vector multiplications in order to avoid expensive matrix-matrix multiplications. 
Hence, one can write
\begin{align}
{\mathbf x}_{\text{TPE, ZF}}= \sum_{l=0}^{J-1}{w_l {\hat{\mathbf x}}_l},
\end{align}
where
\begin{align}\label{iterative-TPE}
{\hat{\mathbf x}}_l =
\Bigg\{
\begin{array}{lr}
        {\mathbf H}^T{\mathbf y},  &  l=0\\
        {\mathbf H}^T({\mathbf H}{\hat{\mathbf x}}_{l-1}) , & 1 \le l \leq J-1.
\end{array}
\end{align}
It can be shown that for given $w_{l}$'s, the calculation of (\ref{iterative-TPE}) 
requires $\mathcal{O}(JKN)$ operations \cite{Izadinasab2022}. However, $w_l$'s depend on the normalization factor $\alpha$, which has a key impact on the convergence of the TPE. It is shown in \cite{Sessler2002} that for the TPE of the ZF solution, the optimal normalization factor $\alpha_{\text{opt}}$ with respect to the asymptotic convergence speed happens when we have
\begin{align}\label{norm_opt}
   \alpha_{\text{opt}} =\frac{2}{\lambda_{min}({\bf G})+\lambda_{max}({\bf G})}.
\end{align}
However, the calculation of the channel Gram matrix $\bf G$ itself requires $\mathcal{O}(NK^2)$ operations. In addition, the calculations of $\lambda_{min}({\bf G})$ and   $\lambda_{max}({\bf G})$ require $\mathcal{O}(K^3)$ operations.
In \cite{Izadinasab2022}, an efficient approximate method for calculating $\alpha_{\text{opt}}$ is proposed for massive MIMO systems such that the overall computational complexity of the TPE-based detector will be $\mathcal{O}(JKN)$.

%%%%%%%%%%%%%%%%%%%%%%%%
%%% 
%%%%%%%%%%%%%%%%%%%%%%%%

\section{The Proposed Learned TPE}\label{The_Proposed_Learned_TPE} 
In this section, we propose exploiting DL for estimating the coefficients of the TPE. In the following, we explain the benefit of optimizing the coefficients over a large number of training examples using DL. To illustrate this, we rewrite ${\bf W}_{\text{ZF}}$ in (\ref{W_ZF}) in terms of  its approximated TPE matrix using $J$ terms, i.e., ${\bf W}_{\text{TPE, ZF}}$, and the higher order TPE terms as follows:  %If we rewrite (\ref{W_ZF}), we have
\begin{align}\label{}
{\bf W}_{\text{ZF}}&= \sum_{l=0}^{J-1}{w_l\Big( {\bf H}^T{\bf H} \Big)^l {\bf H}^T} + \sum_{l=J}^{\infty}{w_l\Big( {\bf H}^T{\bf H} \Big)^l {\bf H}^T}\nonumber\\\label{Two_Summations}
&= {\bf W}_{\text{TPE, ZF}} + \sum_{l=J}^{\infty}{w_l\Big( {\bf H}^T{\bf H} \Big)^l {\bf H}^T}.
\end{align}
The coefficients obtained by inserting (\ref{norm_opt}) in (\ref{w_l's}) are optimal with respect to the asymptotic convergence speed. However, they are not necessarily the best coefficients for the approximation in (\ref{W_ZF}) using a finite value of $J$. Using TPE, the second summation in (\ref{Two_Summations}) is discarded, which may result in a coarse approximation, especially for smaller TPE orders. Therefore, using DL can help optimize the coefficients for a specific value of $J$, resulting in a better approximation. Model-driven DL is a promising candidate for such an approximation since the expansion terms in (\ref{W_ZF}) can be mapped to the DNN with $J$ layers.      

Let $\big\{ {\bf H}^{(m)} \big\}_{m=1}^{M}$ denote the set of training samples of the channel matrix. We define parameters ${\bf \Theta} = \big\{ w_0, \cdots, w_{J-1}  \big\}$ for TPE with the order of $J$. We initialize these parameters using the constant normalization factor proposed in \cite{Izadinasab2022}. The mean square error (MSE) is adopted as the loss function in order to express the distance between matrices ${\bf W}_{\text{ZF}}$ and ${\bf W}_{\text{TPE, ZF}}$, i.e, \begin{align}\label{loss_function}
{\mathcal{L}\big({\bf \Theta}\big)} = \frac{1}{M}\sum_{m=1}^{M}{\vert\vert {{\bf W}_{\text{ZF}} - {\bf W}_{\text{TPE, ZF}}}\vert\vert}_{F}^{2},
\end{align}
where $\vert\vert {\cdot\vert\vert}_{F}$ denotes the Frobenius norm. Note that ${\bf W}_{\text{TPE, ZF}}$ is a function of parameters $w_l$'s. After training, which is performed offline, the trained parameters ${\bf \Theta}^\star = \big\{ w_0^\star, \cdots, w_{J-1}^\star  \big\}$ are used for detecting the transmitted signal vector using (\ref{iterative-TPE}) for new channel samples in the online detection stage.  

We note that by using the loss function in (\ref{loss_function}), the learned parameters can be used for any arbitrary modulation as the training is conducted only using the samples of the channel matrix.   
%%%%%%%%%%%%%%%%%%%%%%%%
%          Complexity Analysis
%%%%%%%%%%%%%%%%%%%%%%%%
\begin {table*}[h]
\caption {Computational Complexity In Terms of Complex Multiplications} \label{Complexity_Table} 
\vspace{-0.5cm}
\begin{center}
% \resizebox{\columnwidth}{0.1\columnwidth}{
\begin{tabular}{ |c|c|c| }
\hline
Detector    &  Complex Multiplications & Complexity Order \\\hline
\hline
ZF/MMSE     &      $\frac{1}{2}NK^2 + \frac{1}{2}K^3 + \frac{3}{2}NK + \frac{5}{2}K^2$                & $\mathcal{O}(NK^2 + K^3)$\\ \hline
TPE-ZF ($\alpha_{\text{constant}}$)\cite{Izadinasab2022}           &      $2JNK - NK + JK$                &  $\mathcal{O}((2J-1)NK)$\\ \hline
TPE-ZF ($\alpha_{\text{power}}$)\cite{Izadinasab2022}           &      $2JNK - NK + JK + \frac{1}{2}KJ(J-1) + 2(K+J)$                &  $\mathcal{O}((2J-1)NK)$\\ \hline 
Proposed           &      $2JNK - NK + JK$                &  $\mathcal{O}((2J-1)NK)$\\ \hline 
\end{tabular}
% }
\end{center}
\end {table*}

\section{Complexity Analysis}\label{Section_Complexity_Analysis}
%%%%%%%%%%%%%%%%%%
% In this section, we analyze the computational complexity of the proposed approach for the problem of signal detection for uplink massive MIMO systems. 
Table \ref{Complexity_Table} shows the computational complexity of several detectors for massive MIMO systems in terms of complex multiplications. The ZF or MMSE detector requires a computational complexity of $\mathcal{O}(NK^2 + K^3)$ as both involve a matrix-matrix multiplication and a matrix inversion. The TPE-based detector in \cite{Sessler2002} has a computational complexity of $\mathcal{O}(NK^2 + K^3 + (2J-1)NK)$ as it requires the calculation of the Gram channel matrix and its extreme eigenvalues for obtaining $\alpha_{\text{opt}}$ in (\ref{norm_opt}). The TPE-based detector in \cite{Izadinasab2022} proposes an approximate solution for calculating $\alpha_{\text{opt}}$ for both small and large loading factors of massive MIMO. For a small loading factor, a constant normalization factor is offered. Hence, the coefficients can be computed at the preprocessing stage. For large loading factors of massive MIMO, an approximate approach using the power method is offered in \cite{Izadinasab2022} with additional computational complexity in the online detection stage. The overall computational complexity of the approach in \cite{Izadinasab2022} for both cases is $\mathcal{O}((2J-1)NK)$. Our proposed learned TPE-based detector also requires a computational complexity of $\mathcal{O}((2J-1)NK)$ if we neglect the computational complexity of the offline training stage where the coefficients are optimized. However, as is shown in Section \ref{Section_Simulation_Results}, our learned TPE-based detector outperforms the detector in \cite{Izadinasab2022} for a given $J$ and can achieve the BER of the ZF or MMSE solution using a smaller $J$. Hence, the computational complexity of the online detection stage of the proposed detector will be reduced at the cost of computational complexity in the offline training stage.     

%%%%%%%%%%%%%%%%%%%%%%%
%          
%%%%%%%%%%%%%%%%%%%%%%%
\section{Simulations Results}\label{Section_Simulation_Results}

We have implemented the proposed approach in Python using the TensorFlow library. 
%%%%%%%%%%%%%%%%%%%%%%%
\subsection{Training Setup}
\begin {table}[t]
\caption {Training Setting} \label{Training_Setting} 
\vspace{-0.6cm}
\begin{center}
\resizebox{\columnwidth}{!}{
\begin{tabular}{ |c|c|c| }
\hline
Parameters    &  \multicolumn{2}{c|}{Values} \\
\hline\hline
$N$    & $128$ & $64$\\
\hline
$J$    & $\{2,3,4\}$ & $\{4,5,6,8\}$\\
\hline 
$K$    & \multicolumn{2}{c|}{$16$}\\
\hline
Size of training data, $M$    & \multicolumn{2}{c|}{$ 10,000$}\\
\hline
Mini-batch size    & \multicolumn{2}{c|}{$ 200$}\\
\hline
Optimization method    &\multicolumn{2}{c|}{Adam} \\
\hline 
Initial learning rate    &\multicolumn{2}{c|}{$0.001$}\\
\hline
Exponential decay rate    &\multicolumn{2}{c|}{$0.9$}\\
\hline
Number of epochs    & \multicolumn{2}{c|}{$2,000$}\\
\hline
\end{tabular}
}
\end{center}
\end {table}

The training setup for the proposed approach is given in Table \ref{Training_Setting}.
The adaptive momentum estimation (Adam) optimizer \cite{Kingma2015Adam} is deployed for minimizing the loss function. We generate $10,000$ samples of Rayleigh fading channel matrices with i.i.d. entries. The starting learning rate for Adam optimizer is set to $\beta_0 = 0.001$. After each epoch, the learning rate decays exponentially such that
\begin{align}\label{exponential_decay}
   \beta_t = \beta_0\times {0.9}^t,  
\end{align}
where $\beta_t$ is the updated learning rate after $t$-th epoch.

%%%%%%%%%%%%%%%%%%%%%%%
\subsection{BER Performance Evaluation} 
In this subsection, we investigate the error performance of the proposed DL-based detector.

\begin{figure}[t]
     \centering
     \includegraphics[width= 0.5\textwidth, height = 5cm, trim = 145mm 56mm 165mm 33mm, clip]{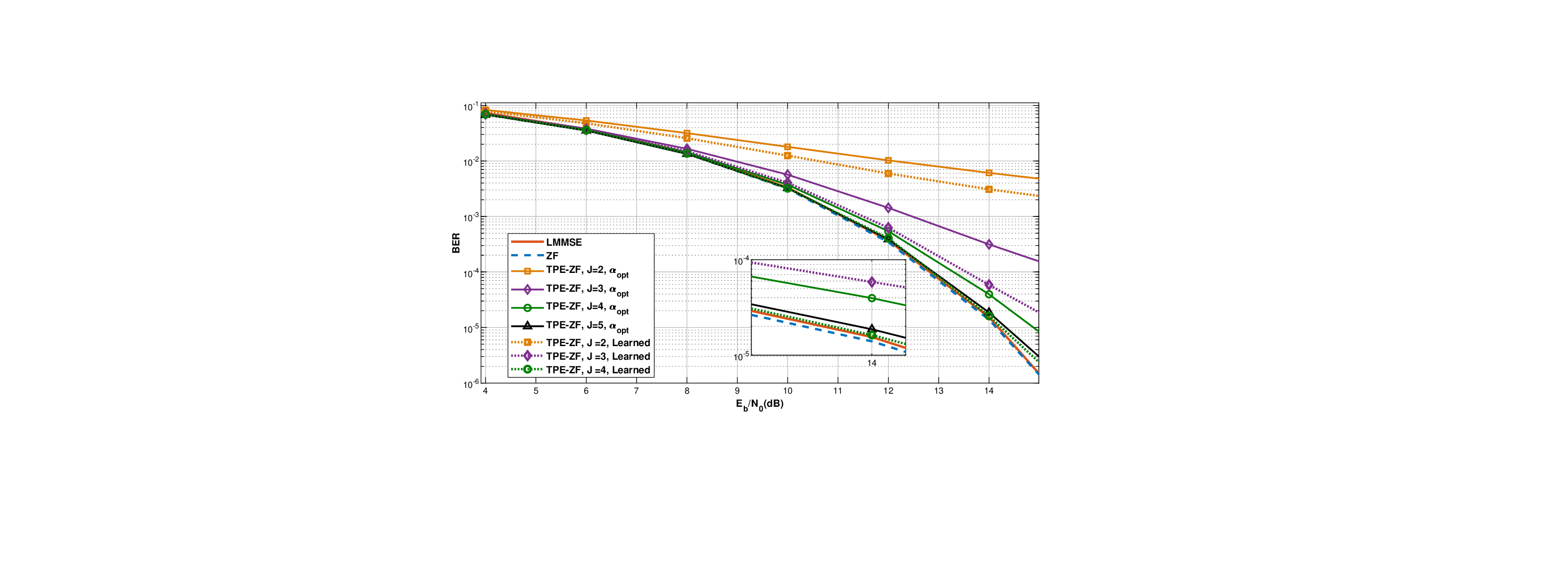}
     \caption{BER performance for a massive MIMO system with $N = 128$, $K = 16$, and $16$-QAM modulation.}
     \label{128x16}
\end{figure}

In Fig. \ref{128x16}, we consider a $128\times 16$ massive MIMO system with $16$-QAM modulation. For this system, the proposed DL-based TPE-ZF detector exhibits a better error performance with the trained coefficients compared to the case when the optimal normalization factor $\alpha_{\text{opt}}$ in (\ref{norm_opt}) proposed in \cite{Sessler2002} are used. The proposed method requires $J=4$ in order to approach the error performance of the ZF or MMSE solution, while for the case with $\alpha_{\text{opt}}$, $J=5$ is required. For this system, which has a quite small loading factor, according to table \ref{Complexity_Table}, the number of complex multiplications in the online detection stage is reduced by $22.21\%$ compared to the TPE-based detector in \cite{Izadinasab2022}, which is a reduced complexity version of the method in \cite{Sessler2002}. Note that in systems with $N \gg K$, ZF and MMSE schemes exhibit the same error performance. 

\begin{figure}[t]
     \centering
     \includegraphics[width= 0.5\textwidth, height = 5cm, trim = 145mm 56mm 165mm 33mm, clip]{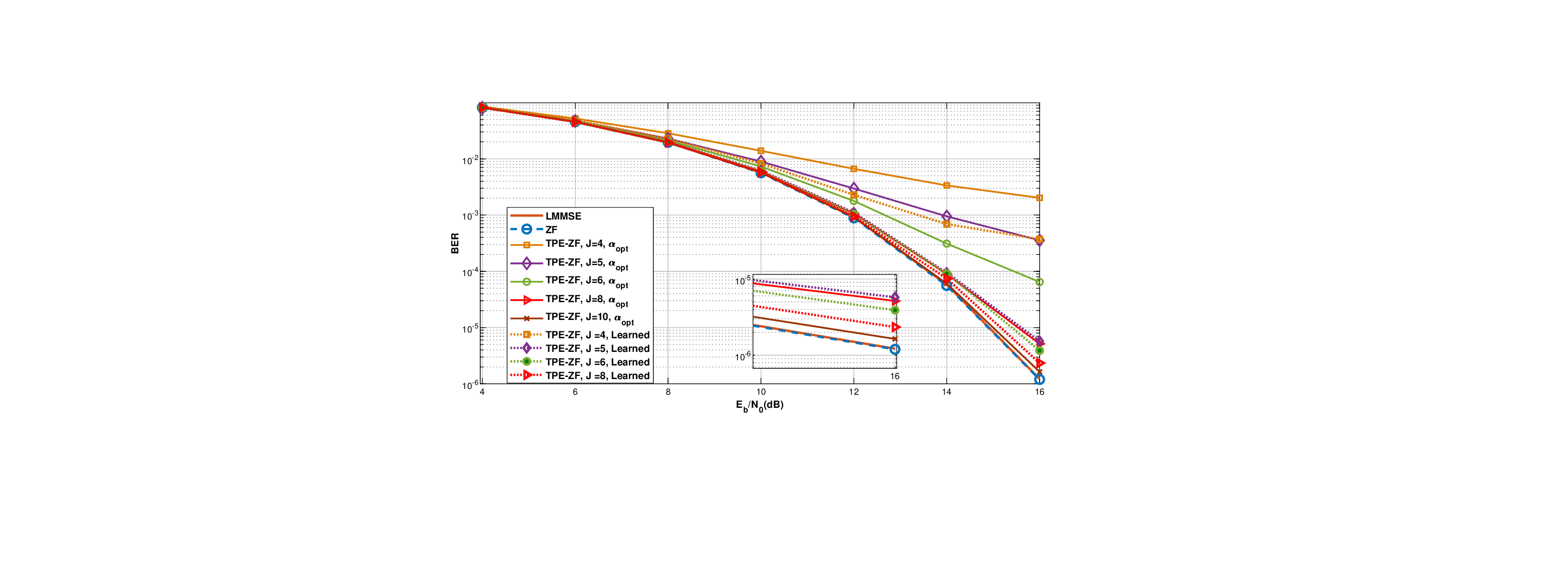}
     \caption{BER performance for a massive MIMO system with $N = 64$, $K = 16$, and $16$-QAM modulation.}
     \label{64x16}
\end{figure}

Fig. \ref{64x16} shows the BER performance for a $64 \times 16$ massive MIMO system with $16$-QAM modulation. For this system, the proposed detector outperforms the TPE-ZF with $\alpha_{\text{opt}}$. In particular, when $\alpha_{\text{opt}}$ is used  
the TPE-ZF needs $J=10$ to approach the ZF or MMSE solution while our proposed detector requires $J=8$. For this system, the computational saving by using the proposed detector is $24.03\%$ compared to the TPE-based detector for systems with large loading factors in \cite{Izadinasab2022}. Note that as was mentioned earlier the optimal coefficients in terms of asymptotic convergence are not necessarily the best ones for error performance.   

%%%%%%%%%%%%%%%%%%%%%%%%
%          Conclusion
%%%%%%%%%%%%%%%%%%%%%%%%
\section{Conclusion}\label{Section_Conclusion}
In this paper, we proposed a TPE-based detector for the uplink of massive MIMO systems. A model-driven DL approach was considered in order to optimize the coefficients of the TPE-based detector. The proposed method outperforms the existing TPE-based detectors at the expense of higher computational complexity in the offline training process while the online detection complexity is reduced.

%%%%%%%%%%%%%%%%%%%%%%%
%          Appendix
%%%%%%%%%%%%%%%%%%%%%%%
% \section*{Appendix A\\ proof of Proposition \ref{Proposition1}}\label{App1}

%%%%%%%%%%%%%%%%%%%%%%%%
%          References
%%%%%%%%%%%%%%%%%%%%%%%%
\bibliographystyle{IEEEtran}
\bibliography{References.bib}
\end{document}